\documentstyle[aps,pra,prabib,amsmath,amssymb,amsfonts,epsfig]{revtex}
\newcommand{\cen}[1]{\begin{center}#1\end{center}}
 
\newcommand{\eqname}[1]{\label{eq:#1}}
\newcommand{\bgar}{\begin{eqnarray}}
\newcommand{\enar}[1]{\label{eq:#1}\end{eqnarray}}
\newcommand{\valass}[1]{\left|#1\right|}

\newcommand{\braket}[2]{ \langle #1 | #2 \rangle }
\newcommand{\braopket}[3]{ \left\langle #1 \right| #2 \left| #3 \right\rangle }
\newcommand{\ketbra}[2]{ | #1 \left\rangle \right\langle #2 |}

\newcommand{\derivexp}[2]{\frac{\partial #1}{\partial #2}}

\newcommand{\el}{{{\bf e}}}
\newcommand{\El}{{{\bf E}}}

\newcommand{\J}{ {\bf J}}

\newcommand{\k}{ {\bf k}}
\newcommand{\kh}{\hat {\bf k}}
\newcommand{\w}{ {\bf w}}
\newcommand{\wh}{\hat {\bf w}}
\newcommand{\vel}{ {\bf v}}
\newcommand{\x}{ {\bf x}}
\newcommand{\zh}{ {\hat {\bf z}}}
\newcommand{\sigh}{ {\hat {\bf \sigma}}}

\newcommand{\xperp}{ {\bf x}_\perp}
\newcommand{\kperp}{ {\bf k}_\perp}
\newcommand{\eps}{\epsilon}
\newcommand{\epsop}{{\hat \epsilon}}
\newcommand{\projop}{{\hat {\mathcal P}}^{\bf k}}

\newcommand{\eq}[1]{(\ref{eq:#1})}
\newcommand{\al}[1]{^{(#1)}}

\twocolumn

\begin{document}
 \title{\bf Slow group velocity and Cherenkov radiation.}
\author{I. Carusotto$^{(a,b,*)}$, M. Artoni$^{(c)}$, G. C. La
Rocca$^{(a,b)}$ and F. Bassani$^{(b)}$}
 \address{
(a) Dipartimento di Fisica ``E. R. Caianiello'', Universit{\`a} di Salerno,
via S. Allende, I-84081 Baronissi (Sa), Italy \\
(b) INFM, Scuola Normale Superiore, Piazza dei Cavalieri 7,
56126 Pisa, Italy \\
(c) INFM, European Laboratory for non-Linear Spectroscopy, Largo E. Fermi
2, 50125 Florence, Italy
 \\[3mm]
\date{\today}
{\rm
\parbox[t]{14.1cm}{\rm
We theoretically study the effect of ultraslow group velocities on the
emission of
Vavilov-Cherenkov radiation in a coherently driven medium.
We show that in this case the aperture of the {\em group cone} on which the
intensity of the radiation
peaks is much smaller than that of the usual {\em wave cone} associated
with the Cherenkov coherence condition.
We show that such a singular behaviour may be observed in a coherently
driven ultracold atomic gas.
 \\[4mm]
 }}}
\maketitle

\vspace{3mm}

The recent observation of ultraslow group velocities in coherently driven
media~\cite{SlowLight,Magnet,Hau}
has opened the way towards new regimes of light
propagation~\cite{LuceFerma} and non--linear optics at very weak
intensities~\cite{Photo}.
In this letter, we investigate the effect of ultraslow group velocities on
the Cherenkov radiation emitted
by a charged particle uniformly moving with a  velocity larger than the
phase velocity of light.
In the case of isotropic and non-dispersive media,
the surface on which the intensity peaks coincides with the well known {\em
wave cone} orthogonal to the wavevector of the emitted light and
the aperture of which is determined by the usual Cherenkov
coherence condition.
In the case of highly dispersive media, as we show here, the intensity is
instead peaked on the surface of a
{\em group cone}~\cite{GroupCone} neither orthogonal to the phase nor to the group velocity
of the emitted light and much narrower
than the wave cone.
First, we develop a general theory for Cherenkov emission in arbitrary
dispersive
and non-isotropic media from which we obtain analytical expressions for the
electric field intensity profile and, in particular, for the {\em group
cone} aperture.
Then, as an example, we investigate the optical properties of a coherently
driven ultracold
atomic cloud of $^{23}\textrm{Na}$ atoms which appears to be amenable
to the observation of such a singular behaviour.
A clear distinction between wave cone and group cone is expected to clarify
the role of the group velocity in
the process of Cherenkov emission.

Consider a charged point-like particle moving with constant velocity
$\w=w\zh=c\beta\zh$ through a non-absorptive
and homogeneous medium characterized by an anisotropic and dispersive
hermitian dielectric tensor $\hat{\eps}(\omega)=\eps_{i,j}(\omega)$.
The effects of a weak absorption will be considered afterward for the case
of interest.
Using Maxwell equations in Fourier space, we write the $(\k,\omega)$
component of the radiated electromagnetic field $\El(\k,\omega)$ in terms
of the corresponding Fourier components of the current density
${\J}(\k,\omega)=2\pi q\,\delta(\omega-\k\cdot \w)\,\w$ as
\begin{equation}\eqname{Electric}
\El(\k,\omega)=\frac{4\pi i \omega}{c^2}\left(k^2
\projop-\frac{\omega^2}{c^2}\epsop(\omega)\right)^{-1}\J(\k,\omega)
\end{equation}
where $\projop_{i,j}=\delta_{i,j}-\frac{k_i k_j}{k^2}$ is the projection
operator onto the subspace orthogonal to $\k$ and $k^2=\sum_i k_i k_i$ is
the square modulus of $\k$.

The poles of \eq{Electric} determine the propagating modes of the
electromagnetic field through the well-known Fresnel
equation~\cite{LandauECM}
\begin{equation}\eqname{PropModes}
\left(k^2
\projop-\frac{\omega_\alpha^2}{c^2}\epsop(\omega_{\alpha})\right)\el\al{\alpha}=
0.
\end{equation}
For each wavevector $\k$, the different modes are characterized by the
frequency $\omega_{\alpha}$ and
the polarization unit vector $\el\al{\alpha}$ normalized as
$\braket{\el\al{\alpha}}{\el\al{\alpha}}=\sum_i \el^{(\alpha)*}_i
\el\al{\alpha}_i=1$.
We will not account here for longitudinal modes,
assuming $\omega^{2} \neq 0$ and $\textrm{det}[\epsop(\omega)]\neq 0$ for
all values of $\omega$.
For each mode, the group velocity can be shown to be~\cite{footnote4}
\begin{equation}\eqname{GroupVeloc}
\vel\al{\alpha}_g=\nabla_{\k}\omega_{\alpha}=c^2\frac{
2\k-\el^{(\alpha)}\braket{\el\al{\alpha}}
{\k}-\el^{(\alpha)*}\braket{\k}{\el\al{\alpha}} }{
\braopket{\el\al{\alpha}}{\frac{\partial}{\partial\omega}
 \left(\omega^2 \epsop(\omega) \right)}{\el\al{\alpha}}  }.
\end{equation}
We shall also restrict our attention to the simple case of a medium with
rotational symmetry around
the direction of the charge velocity; in this case, the revolution symmetry
of the dispersion surface
$\omega_{\alpha}(\k)=\omega$ guarantees the parallelism of the components
of the group velocity and the
wavevector perpendicular to the $\zh$ axis.

The electric field at a position $\x=(\xperp,z)=(x_\perp {\hat {\bf
u}}_\perp,z)$ can be obtained from the inverse Fourier transform of
\eq{Electric}.
At sufficiently large distances from the charge trajectory, only the poles
of \eq{Electric} effectively contribute to the Fourier transform.
In fact, the electric field at these distances is given by the resonant
excitation of propagating modes, while the non-resonant contribution from
all other modes decays out in space with a faster power
law~\cite{footnote1}.
Within such an approximation we can write for the $i^{th}$ component,

\begin{eqnarray}
E_i(\xperp,z,t)&=&\frac{2i q}{c^2}
\int_0^\infty\!d\omega\sum_{\alpha=1,2}
\frac{k_\perp\al{\alpha}\omega\braket{\el\al{\alpha}}{\hat{\w}}}{\mu 
_\alpha}
\nonumber\\
&\times &
 \sqrt{\frac{i}{2\pi k_\perp\al{\alpha}x_\perp}}e^{i
k_\perp\al{\alpha}
x_\perp}e^{i\frac{\omega}{w}(z-w t)}\,\el_{i}\al{\alpha}.
\eqname{ElectricXT}
\end{eqnarray}
Here, $\wh=\w/w$ is the direction of the charge velocity $\w$ and
$\mu _\alpha=\braopket{\el\al{\alpha}}
{\frac{\partial}{\partial k_\perp}\left(k^2\projop\right)}{\el\al{\alpha}}$
is a weight factor.
In obtaining the expression \eq{ElectricXT} the angular integration in
$\kperp$ has been already
performed in the large $k_\perp x_\perp$ limit so that, for each position
 $\xperp$, only the field components with $\kperp$ parallel to $\xperp$ have a
  stationary phase and therefore give a nonvanishing contribution to the
integral.

\begin{figure}
\cen{\psfig{figure=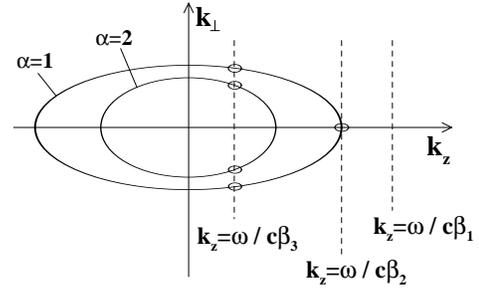,height=1.5in}}
\caption{Schematic plot of the longitudinal cross section of the dispersion
surfaces for the two propagating $\alpha=1,2$ modes at a given
$\omega$. The vertical lines are the cross sections of the
$k_z=\omega/\beta c$ planes.
For $\beta=\beta_1$, no Cherenkov radiation occurs at $\omega$;
$\beta=\beta_2$ is the thershold velocity for the $\alpha=1$ mode; for
$\beta=\beta_3$, both modes are excited.
\label{fig:DispSurf}}
\end{figure}

Depending on the velocity of the charged particle, for each direction
${\hat {\bf u}}_\perp$ and frequency $\omega$, at most two distinct poles
at $k_\perp\al{\alpha}$ ($\alpha=1,2$) contribute to the integral, which
correspond to propagating modes with wavevector
$\k\al{\alpha}=(k_\perp\al{\alpha}{\hat {\bf u}}_\perp,\omega/w)$ and polarization
$\el\al{\alpha}$.
As shown in fig.\ref{fig:DispSurf}, at a given frequency $\omega$ Cherenkov
light is emitted into the mode $\alpha$ only if the $k_z=\omega/c\beta$
plane has a non-vanishing intersection with the dispersion surface
$\omega_{\alpha}(\k)=\omega$ of the mode.
In the case of a medium with rotational symmetry, this condition is
satisfied if the particle velocity exceeds
 the phase velocity of mode $\alpha$ for $\k$ parallel to $\w$, i.e.
\begin{equation}\eqname{MinVel}
\beta^2>\left.\frac{\omega^2}{c^2k^{(\alpha)2}}\right|_{\k\parallel\w}=\frac{1}{
\eps_\perp\al{\alpha}(\omega)}
\end{equation}
 and the intersection is the $k_\perp=k_\perp\al{\alpha}$ circle.
Here $\eps_\perp\al{\alpha}(\omega)$ are the eigenvalues of the dielectric
constant in the plane perpendicular to the $\zh$ axis.
For an isotropic medium, the condition \eq{MinVel} reduces to the usual
Cherenkov condition $\beta^2\eps(\omega)>1$
~\cite{Jelley}.

The theory developed up to now has considered the case of a non-absorptive
medium for which
the poles $k_\perp\al{\alpha}$ are real; the effect of a weak absorption
consists in the introduction of a small and positive imaginary part
$\textrm{Im}[k_\perp\al{\alpha}]$ in the argument of the exponential in
\eq{ElectricXT} without modifying the pole structure of the integral.
The resulting damping factor $e^{-\textrm{Im}[k_\perp\al{\alpha}]x_\perp}$
accounts for absorption of the emitted radiation during propagation.

\begin{figure}
\cen{\psfig{figure=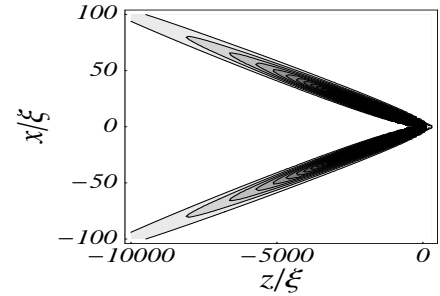,width=2.4in,height=1.65in}}
\caption{Intensity contour--plot of the longitudinal cross-section of the
{\em group cone}
at $t=0$ for $v_g^{\perp}/w=0.01$.
\label{fig:ElProf}}
\end{figure}

\begin{figure}
\cen{\psfig{figure=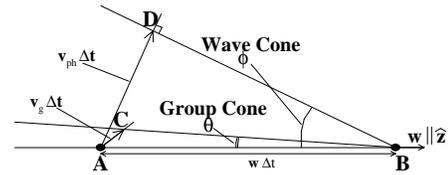,width=2.3in}}
\caption{Geometrical construction for the group cone.
During the time $\Delta t$, the charge moves from $A$ to $B$ with
$\vec{AB}=\w\,\Delta t$,
while the radiation emitted in A propagates from $A$ to $C$ with
$\vec{AC}=\vel_g\,\Delta t$.
As discussed in the text, the straight line joining B and C is a
generatrix of the {\em group cone}.
\label{fig:GeomConstr}}
\end{figure}

We now consider a medium with a narrow transparency window centered at
${\bar \omega}$;
for a single polarization state $\alpha=1$,
the absorption factor in the neighborhood of ${\bar \omega}$ can be
approximated by
$\textrm{Im}[k_\perp\al{1}]\simeq\frac{\eta}{2}(\omega-{\bar \omega})^2$ with
$\eta={\partial^2\textrm{Im}[k_\perp\al{1}]}/{\partial \omega^2}$.
Inserting this expression into \eq{ElectricXT}, we finally obtain an
explicit expression for
the electric field intensity profile
\begin{equation}\eqname{GroupCone}
\valass{\El(\x,t)}^2=\frac{Aq^2}{x_\perp^2}\exp\left[-\frac{1}{\eta
x_\perp}\left(\frac{z}{w}+\frac{x_\perp}{v_r}-t\right)^2\right]
\valass{\braket{\el\al{1}}{\wh}}^2,
\end{equation}
where $\el\al{1}$ is the polarization vector of the mode at frequency
${\bar \omega}$ and
$A=4k_\perp\al{1}{\bar \omega}^2 / c^4\mu _1^2\eta$.
The radial velocity $v_r$ is defined according to
\begin{equation}\eqname{vr}
v_r^{-1}=\left[\left.\derivexp{k_\perp}{\omega}\right|_{k_z=\frac{{\bar
\omega}}{w}}\right]=\frac{w-v_g^\parallel}{w v_g^{\perp}}
\end{equation}
where $v_g^{\perp}$ and $v_g^\parallel$ are respectively the perpendicular
and parallel components of
the group velocity $\vel_g\al{1}$ with respect to the direction $\wh=\w/w$
of the charge velocity.
As it can be observed in fig.\ref{fig:ElProf}, sufficiently far from the charge, i.e. at points $x_\perp\gg \xi=\eta
v_r^2$, the electric field intensity \eq{GroupCone} is peaked on the {\em
group cone} described by
\begin{equation}\eqname{GroupCone2}
\frac{x_\perp}{v_r}+\frac{z}{w}=t
\end{equation}
whose aperture $\theta$ is equal to
\begin{equation}
\eqname{theta}
\tan \theta=v_r/w.
\end{equation}
In general, $\theta$ is different from the aperture $\phi$ of the {\em wave
cone} orthogonal to the wavevectors of the emitted radiation, 
which is instead given by
$\tan \phi=\bar \omega / w k_\perp\al{1}$.
A simple physical interpretation of the group cone can be put forward in
terms of group velocity~\cite{GroupCone}
considering that, for each direction around the charge velocity, the burst
of Cherenkov light is emitted into a group
of modes centered at $\k\al{1}$
while the peak of the pulse moves in space with a velocity 
equal to the group velocity $\vel_g\al{1}$
experiencing an almost negligible absorption.
The cone defined by this geometrical construction
(fig.\ref{fig:GeomConstr}) can be proven in all cases to be equivalent
to the group cone defined by \eq{GroupCone2} and to coincide,
for the case of an isotropic medium,
with the group cone introduced by Frank~\cite{GroupCone}.
Notice that this cone turns out to be in general neither orthogonal to the
group velocity nor to the wavevector.
In the case of isotropic and non-dispersive media, in which
$\vel_g=\vel_{ph}=\frac{c}{n}\kh$, $n$ being the refractive index, the
group and the wave cones coincide and have an aperture $\phi$ defined by
the usual Cherenkov condition $\beta n \sin\phi=1$.

The weak dispersion of common dielectrics makes the difference between the
group and the wave cones very small and has prevented upto now its
experimental observation~\cite{GroupCone}.
When the group velocity is much smaller than the phase velocity the group
cone is expected to be well separated from the wave cone
(fig.\ref{fig:GeomConstr}) and to have an extremely narrow shape $\theta\ll 
1$.
Following the recent observations of ultraslow light in coherently driven
hot~\cite{SlowLight,Magnet}
and cold~\cite{Hau} atomic gases, such media appear as promising candidates
for the experimental
characterization of the role of group velocity in Cherenkov radiation.

\begin{figure}
\cen{\psfig{figure=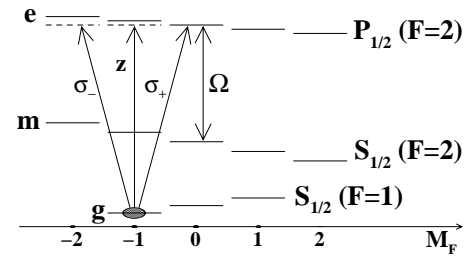,height=1.4in}}
\caption{Scheme of the $^{23}\textrm{Na}$ atomic levels involved in the
optical process
under examination.
\label{fig:LevelScheme}}
\end{figure}

Consider a cloud of ultracold atoms~\cite{footnote5} in a three-level
$\Lambda$-type configuration and take, as a specific example, the case of
$^{23}\textrm{Na}$ atoms (fig.\ref{fig:LevelScheme}) magnetically trapped
in the $M_F=-1$ sublevel of the $F_g=1$ ground state. Let $\omega_g$ be the
frequency of such a state.
The other hyperfine component of the $\textrm{S}_{1/2}$ ground state is a
metastable state with $F_m=2$
approximately $1.8\,\textrm{GHz}$ blue-detuned with respect to the ground
state.
A weak coherent field polarized along the trap magnetic field drives the
optical transition from the metastable state to the $F_e=2$ hyperfine
component of the $\textrm{P}_{1/2}$
excited state.
Its Rabi frequency $\Omega$ is smaller than the excited state linewidth
$\gamma_e\simeq2\pi\cdot 10\,\textrm{MHz}$ while its frequency is taken to
be exactly on resonance with the
transition between the $M_F=0$ Zeeman components
$\omega_{dr}=\omega_e[M_F=0]-\omega_m[M_F=0]$.
Assuming the charge velocity to be parallel to the trap magnetic field, the
rotational symmetry of the system around the $\zh$ axis
implies~\cite{Cornwell} the following decomposition of the dielectric
tensor
$\epsop(\omega)=\eps_z(\omega)\ketbra{\zh}{\zh}+\eps_+
(\omega)\ketbra{\sigh_+}{\sigh_+}+\eps_-(\omega)\ketbra{\sigh_-}{\sigh_-}$.
For each frequency $\omega$ and direction $\kh$, the two propagating modes
defined by \eq{PropModes} are generally non-degenerate except at high
symmetry points and have elliptical polarizations; from the point of view
of the spatial symmetry of the optical constants, the polarized atomic
cloud is in fact not only uniaxial, but also optically active.
Since the linewidth $\gamma_m$ of the metastable $m$ state is orders of
magnitude smaller~\cite{Hau} than that of the excited $e$ state $\gamma_e$,
electromagnetically induced transparency (EIT) occurs on the $\sigh_+$
polarization in a narrow frequency window of linewidth
$\Gamma=\Omega^2/\gamma_e\ll \gamma_e$ around
$\omega_+=\omega_e[M_F=0]-\omega_g$~\cite{Arimondo} where absorption is
quenched and dispersion enhanced so as to give slow light propagation.
In the same frequency window $\omega\simeq\omega_+$, the transitions from
the ground state to the $M_F=-2,-1$
sublevels of the excited state are sufficiently off-resonance
($\Delta\omega_{z,-}=\omega_e[M_F=-1,-2]-\omega_e[M_F=0]\gg \gamma_e$) so as
to give a positive and relatively frequency-flat background contribution to
the $\sigh_-$ and $\zh$ components of the dielectric tensor
\begin{eqnarray}
\eps_{+}&=&1+\frac{4\pi f_+}{\omega_+-i\gamma_e-\omega-
\frac{\valass{\Omega}^2}
{\omega_+-i\gamma_m-\omega}} \\
\label{eq:EIT}
\eps_{z,-}&=&1+4\pi\chi^{\infty}_{z,-}=1+\frac{4\pi
f_{z,-}}{\Delta\omega_{z,-}}.
\label{eq:Backgr}
\end{eqnarray}
The oscillator strengths $f_{\pm ,z}$ are proportional to the atomic
density times the square of the dipole moment of the optical transition and
differ from each other depending on the relevant Clebsch-Gordan
coefficients. 
The detunings $\Delta\omega_{z,-}$ follow from the Zeeman splitting of the
atomic levels, which implies that the background refraction can be
experimentally controlled by tuning the magnetic field;
with $^{23}\textrm{Na}$ atoms, a splitting $\Delta\omega_{-}$ of about
$2\pi\cdot 40\,\textrm{MHz}$ occurs in a reasonable magnetic field of the
order of $80\,\textrm{G}$.
Given the different magnetic moments of metastable and excited states, the
weak dressing field can not effectively dress the optical transitions other
than the resonant one between the $M_F=0$ sublevels.
Thanks to their large detuning with respect to $\omega_+$, all transitions
involving the other hyperfine components of the excited state can be
neglected.

The threshold velocity \eq{MinVel} for Cherenkov emission at $\omega_+$ is
determined by the background refractive index \eq{Backgr}.
If the velocity is larger than the threshold value, light is radiated into
a mode with non-vanishing $k_\perp\al{1}$; in presence of EIT, such a
radiation will propagate with ultraslow group velocity without being
absorbed.
The magnitude of the group velocity \eq{GroupVeloc} is mainly determined by
the dispersive properties of EIT while its
direction depends on the background refractive index only.
Group velocities as low as $17\,\textrm{m/sec}$ have been reported in an
ultracold sodium gas~\cite{Hau} and this means that
 the group cone would have an extremely narrow shape, practically a
cylindrical one ($\theta \ll 1$).
From the parameters of~\cite{Hau}, the characteristic length $\xi=\eta
v_r^2\simeq \eta v_g^{\perp\,2}$ turns out of
the order of $10\,\mu \textrm{m}$ and the background susceptibility
$4\pi\chi ^\infty_{z,-}$ is of the order of $10^{-2}$,
 i.e. much larger than that usually found in gas Cherenkov
counters~\cite{Jelley}.

The theory we have developed is based on the assumption of a homogeneous
medium.
This approximation is reasonable provided the overall size of the atomic
cloud is much larger than the wavelength
 of the Cherenkov light whose detection should be performed within the
cloud itself so to avoid reflection
 effects at the edges of the cloud.
For the case of EIT media, a picture of the group cone can be taken
exploiting the very large cross section for
resonant two-photon absorption processes~\cite{16}: the absorption
coefficient experienced by a laser field on resonance with another optical
transition starting from the $M_F=0$ sublevel of the metastable $m$ state
is in fact proportional to the local intensity of the Cherenkov radiation
at $\omega_+$ which forms the narrow group cone~\cite{Photo}.
Since the interaction of the charge with the atoms of the cloud results not
only on Cherenkov emission but also on other heating and ionization
processes~\cite{LandauECM}, it is necessary to reduce the importance of
such short-range processes by making the charge travel in a region of space
free from atoms.
For the case of an atomic cloud, this can be achieved e.g. by using the
repulsive potential of a blue-detuned laser so as to create a sort of
``tunnel'' through the cloud; a small cylindrical hole with a radius of the
order of the wavelength does reduce the yield of the Cherenkov radiation,
but does not affect the qualitative features of the Cherenkov pulse
propagating in the surrounding medium~\cite{Tunnel}.
Unfortunately, the intensity of the Cherenkov radiation emitted by a single
electron is rather low. For the $^{23}\textrm{Na}$ parameters and
statistically independent electrons, a photon is emitted in the mode under
consideration each $10^7$ electrons.
This problem may be overcome by looking at the electric field generated by
a very large number of
electrons at a time: since the radial velocity $v_r$ is much smaller than
the charge velocity $w$,
the profile of the group cone would not be smeared out even if the spatial
extension of the bunch of electrons
is much longer than the wavelength of the emitted light.

In summary, we have developed a general theory for Cherenkov emission in
arbitrary
non-isotropic and dispersive dielectrics and we have given an analytic
expression for the {\em group cone} over which the intensity of
the emitted light is maximum.
Unlike in isotropic and non-dispersive media,
the group cone is here much narrower than the {\em wave cone} defined
by the usual Cherenkov coherence condition and is neither orthogonal
 to the phase nor the group velocity.
This conceptual distinction becomes of great physical relevance in media
exhibiting slow
light propagation.
For the realistic case of a coherently driven ultracold $^{23}\textrm{Na}$
gas, the geometrical and dispersive properties of the dielectric tensor
are shown to be favourable to the experimental characterization of
the role of the group velocity in the process of Cherenkov emission.

I. C. and G.L. thank F. Illuminati and S. de Siena for enlightening
discussions.

\begin{references}
\bibitem{footnote5}
In usual EIT experiments the effect of Doppler broadening in hot gases is
overcome by choosing a dressing
field co-propagating with the probe field~\cite{Arimondo}; unfortunately,
the conical geometry of Cherenkov
emission does not allow for such a choice, so that one is
forced to use an ultracold sample
in which Doppler broadening is absent.
\end{references}
\end{document}